\documentclass[prx,floatfix,twocolumn,tightenlines,nofootinbib,superscriptaddress]{revtex4-1}
\usepackage{amsmath,amsthm,amssymb}
\usepackage{graphicx}
\usepackage{times}
\usepackage{hyperref}
\hypersetup{
colorlinks=true,
linkcolor=red,
citecolor=red,}
\usepackage{textcomp}
\usepackage{xcolor}
\usepackage{enumerate}
\usepackage{multirow}
\usepackage{tabularx}
\usepackage{dcolumn}
\usepackage[mathscr]{euscript}
\usepackage{mathtools}
\usepackage{textcomp}
\usepackage{calrsfs}
\usepackage[separate-uncertainty=true]{siunitx}
\usepackage{soul}
\usepackage{comment}
\usepackage{dsfont}
\usepackage{amsfonts}
\usepackage[normalem]{ulem}
\newtheorem{definition}{Definition}
\newtheorem{prop}{Proposition}
\newtheorem{lemma}{Lemma}
\newtheorem{theorem}{Theorem}

\newtheorem*{definition*}{Definition}

\newcommand\dmu[2]{{\int_{#1}}{d #2}}

\newcommand\s[1]{_{\rm #1}}

\newcommand{\ketbra}[1]{ | #1 \rangle\!\langle #1 |}

\newcommand{\card} {\operatorname{card}}
\newcommand{\rank} {\operatorname{rank}}

\newcommand{\inprod}[2] {{\langle} {#1} {,} {#2} {\rangle}}

\newcommand{\one}{\leavevmode\hbox{\small1\normalsize\kern-.33em1}}

\newcommand{\R}{\mathbb{R}}
\newcommand{\Z}{\mathbb{Z}}
\newcommand{\id}{\mathcal{I}}

\newcommand{\M}{{\cal M}}
\newcommand{\T}{{\cal T}}
\newcommand{\PP}{{\cal P}}
\newcommand{\Q}{{\cal Q}}
\newcommand{\CC}{{\cal C}}
\newcommand{\HH}{{\mathscr H}}
\newcommand{\LL}{{\cal L}}
\newcommand{\EE}{{\cal E}}
\newcommand{\St}{{\cal S}}
\newcommand{\V}{{\cal V}}

\newcommand{\D}{{\cal D}}
\newcommand{\F}{{\cal F}}
\newcommand{\A}{{\cal A}}
\newcommand{\Y}{{\cal Y}}

\newcommand{\MM}{\mathsf{M}}
\newcommand{\pp}{\mathsf{P}}

\newcommand{\trs}{^\mathsf{T}}

\begin{document}
\title{Contextuality of General Probabilistic Theories}
 \author{Farid Shahandeh}
 \email{Electronic address: shahandeh.f@gmail.com}
 \affiliation{Department of Physics, Swansea University, Singleton Park, Swansea SA2 8PP, United Kingdom}


\begin{abstract}
Generalized contextuality refers to our inability of explaining measurement statistics using a context-independent probabilistic and ontological model.
On the other hand, measurement statistics can also be modeled using the framework of general probabilistic theories (GPTs).
Here, starting from a construction of GPTs based on a Gleason-type theorem, we fully characterize these structures with respect to their permission and rejection of generalized (non)contextual ontological models.
It follows that in any GPT construction the three insistence of (i) the no-restriction hypothesis, (ii) the ontological noncontextuality, and (iii) multiple nonrefinable measurements for any fixed number of outcomes are incompatible.
Hence, any GPT satisfying the no-restriction hypothesis is ontologically noncontextual if and only if it is simplicial.
We give a detailed discussion of GPTs for which the no-restriction hypothesis is violated, and show that they can always be considered as subtheories (subGPTs) of GPTs satisfying the hypothesis.
It is shown that subGPTs are ontologically noncontextual if and only if they are subtheories of simplicial GPTs of the same dimensionality.
Finally, we establish as a corollary the necessary and sufficient condition for a single resourceful measurement or state to promote an ontologically noncontextual (i.e. classical) general probabilistic theory to an ontologically contextual (i.e. nonclassical) one under the no-restriction hypothesis.
\end{abstract}

\maketitle

\section{Introduction}

A promise of classical theories is to give a probabilistic account for the statistics resulted from prepare-and-measure experiments on single physical systems without relying upon their particular operational procedure, or {\it context}.
Contextuality~\cite{KS,Cabello2010,Abramsky2011,Acin2015}, and more recently, generalized contextuality~\cite{Spekkens2005}, refer to no-go theorems dismissing context-independent classical models for measured statistics.
Hence, a plausible definition of nonclassicality of statistics is through their generalized contextuality~\cite{Spekkens2008,Schmid2018}.

Besides probabilistic (or {\it ontological}) models, statistics from operational procedures can also be modeled using the framework of {\it general probabilistic theories} (GPTs), an example of which is quantum theory~\cite{Hardy2001,Barret2007,Chiribella2010,Chiribella2011,Hardy2011,Masanes2011,Janotta2014}.
Nonclassical data thus mean that none of their potential GPT explanations can be translated into a noncontextual ontological model~\cite{Spekkens2005}.
Recently, Kunjwal and Spekkens~\cite{Kunjwal2015} and Schmid {\it et~al}~\cite{Schmid2018PRA} provided noncontextuality inequalities for detection of possible nonclassical statistics.
In this Letter, we determine which GPTs for physical phenomena allow and which ones disallow noncontextual ontological models. 
More specifically, we prove a Gleason-type theorem for construction of GPTs and show that, in finite dimensions, any GPT that satisfies the no-restriction hypothesis~\cite{Chiribella2010} must possess simplex sets of nonrefinable effects and states to be ontologically noncontextual.
We then discuss the scenario in which the no-restriction hypothesis is removed.

The practical significance of our analysis lies within a markedly interesting context.
Referring to the scenario of quantum computations with Clifford circuits, stabilizer input states, and Pauli measurements, it is well-known that such computations authorize a classical model making them classically efficiently simulatable~\cite{Mari2012,Veitch2013,Veitch2014,Raussendorf2019}.
This possibility is removed by providing only {\it a single} suitable nonstabilizer input state (or measurement) that ``magically'' enables fault-tolerant universal quantum computing~\cite{Gottesman1999,Bravyi2005,Knill2005,Howard2014,Raussendorf2019}.
Thus, there are scenarios in which only a single extra preparation or measurement procedure is {\it resourceful} in that it simultaneously gives rise to two phenomena.
First, it generates data that render a classical model impossible.
Second, it causes a significant improvement in the performance of some information processing protocols.
Our analysis here is motivated by the first phenomenon which is a prerequisite of the second.

The study of resources for information processing purposes commonly begins with assuming an underlying theory.
Thinking of quantum theory, this is beautifully done within the formalism of quantum resource theories~\cite{Chitambar2018} such as entanglement~\cite{Horodecki2009}, athermality~\cite{Brandao2013,Horodecki2013,Brandao2015,Faist2015,Lostaglio2015,Gour2015}, coherence~\cite{Aberg2006,Baumgratz2014,Winter2016,Streltsov2017}, asymmetry~\cite{Gour2008,Gour2009,Piani2016}, non-Markovianity~\cite{Rivas2014}, and dynamical correlations~\cite{Rivas2015}.
Here, we instead relax the assumption of a specific underlying theory by adopting the generic formalism of GPTs and use our first result to obtain the necessary and sufficient criterion to distinguish between classicality and nonclassicality of a single resourceful measurement or state leading to potential nonclassical information processing advantages~\cite{Schmid2018}.

\section{Ontological Models}

Throughout this paper, we are only interested in the prepare-and-measure experiments on single systems that can be described in finite dimensions.
Operationally, the primitive elements of our physical description in any such experiment are the {\it laboratory prescriptions} for preparations and measurements forming the collections $\PP{:=}\{\pp_k\}$ and $\M{:=}\{\MM_j\}$, respectively~\cite{Spekkens2005}.
The fundamental goal in theoretical physics is to establish assignments between these elements and mathematical objects endowed with a set of rules to determine the outcome probabilities in each measurement.
A first generic step is to meaningfully assign (not necessarily scalar) ``sizes'', called {\it measures}, to measurement outcomes.
Given the finite set $\Omega$ of all outcomes, a set $\omega$ of its subsets on which such assignments are well-defined is called the $\sigma$-algebra of events and the pair $(\Omega{,}\omega)$ is named a {\it measurable space};
see Appendix~\ref{app:revmeas} for a brief review.

Ontological models go beyond the minimal promise of theoretical physics by assuming an underlying {\it ontic} variable space $\Upsilon$ and assigning physical phenomena to {\it elements of reality}.
In this process, the operational elements, i.e. preparations and measurements, correspond most generally to probabilistic preparations and measurements of the ontic variable and thus, they are represented by probability distributions and indicator functions over $\Upsilon$, respectively.
More precisely, one defines also a $\sigma$-algebra $\upsilon$ on $\Upsilon$ and designates ``sizes'' to members of both $\omega$ and $\upsilon$ that are probabilities, aka {\it probability measures}.
Given the collections $\Y$ and $\Q$ of all probability measures on $(\Upsilon{,}\upsilon)$ and $(\Omega{,}\omega)$, respectively, the ontological model then hypothesizes the existence of {\it convex linear} maps $\mu{:}\PP{\to}\Y$ and $\xi{:}\M{\to}\Q$ that assign the \textit{ontic state} $\mu_\pp$ to the preparation procedure $\pp$ and the \textit{ontic measurement} $\xi_{\MM}$ to the measurement procedure $\MM$~\cite{Spekkens2005}.
Thus, for each preparation,
\begin{equation}\label{eq:onticstate}
	\mu_\pp:\upsilon\to[0,1] \quad\text{and}\quad \dmu{\Upsilon}{\lambda\mu_\pp(\lambda)}=1,
\end{equation}
and for each measurement,
\begin{equation}\label{eq:onticmeasure}
	\xi_\MM : \omega{\times}\Upsilon \to [0{,}1] \quad\text{and}\quad \xi_{\MM}(\Omega{|}\lambda)=1 \quad \forall\lambda\in\Upsilon.
\end{equation}
Then, the probability of a particular event $X$ in a measurement $\MM$ given the preparation $\pp$ can be obtained via Bayes' rule,
\begin{equation}\label{eq:onticBayes}
	p(X{|}\pp{,}\MM) {=} \dmu{\Upsilon}{\lambda} \mu_\pp(\lambda) \xi_{\MM}(X{|}\lambda).
\end{equation}

\section{General Probabilistic Theories}

A second approach to the abstraction of operational scenarios is known as {\it general probabilistic theories} (GPTs).
Their constructions begin with assuming a vector space $\V$ whose elements can (at least partially) be ordered and on which an inner-product $\inprod{\cdot}{\cdot}$ can be defined~\cite{Hardy2001,Barret2007,Chiribella2010,Chiribella2011,Hardy2011,Masanes2011,Janotta2014}.
In the example of quantum theory, this vector space is the Banach space $\LL(\HH)$ of all bounded linear Hermitian operators on a Hilbert space $\HH$ with the usual Hilbert-Schmidt inner product.
In contrast to ontological models, in this case, to each measurement event $X{\in}\omega$ we assign a {\it vector} as its ``size'';  cf. Appendix~\ref{app:revmeas}.
Hence, instead of a probability measure we have a {\it probability vector-valued measure} (PVVM) which is a function $E{:}\omega{\to}\V$ satisfying (i) $E(X){\geqslant} 0$ for all $X{\in}\omega$, (ii) $E(\Omega){=}U$ for a fixed nonzero element $U{\in}\V$ called the {\it unit} element, and (iii) $E({\cup_i} X_i){=}{\sum_i} E(X_i)$ for all sequences of disjoint events $X_i{\in}\omega$.
An easy way to make sense of these conditions is by comparing a PVVM to a probability measure where after the substitutions $\V{\mapsto}\R$ and $U{\mapsto}1$ the former simply reduces to the latter; see Appendix~\ref{app:revmeas}.
Each vector $E(X_i)$ is called an {\it effect} where their collection is denoted by $\EE$.
The familiar quantum counterpart of a PVVM is a {\it positive operator-valued measure} (POVM)~\cite{HolevoBook,BuschBook}.

Suppose that along with a measurement having $E(X)$ as an effect the experimenter tosses a (biased) coin with probability $p$ for heads and accepts the occurrences of $X$ only if the coin is heads.
Assuming that the product rule of probability holds, the effect corresponding to the accepted events is given by $pE(X)$.
Hence, it is reasonable to assume that given any operationally legitimate effect $E(X)$, the effect $pE(X)$ for any real number $p{\in}[0,1]$ is also allowed.
Further, suppose that $E_1$ and $E_2$ are two PVVMs and our experimenter performs $E_1$ if the coin is heads and $E_2$ otherwise.
The resulting PVVM is thus $E_p{:=}pE_1 {+} (1{-}p)E_2$ implying that the set of all effects $\EE$ is convex. 
In order that the latter to be a well-defined summation $E_1$ and $E_2$ have to possess a common domain.
Consequently, throughout this paper we assume that the event space $\omega$ is fixed.
Finally, we assume that $\EE$ spans $\V$.
These allow us to state a Gleason-type theorem for GPTs as follows.
\begin{theorem}\label{th:Gleason}
 Any generalized probability measure $q{:}\EE{\to}[0,1]$ satisfying (i) $q(E(X)){\geqslant} 0$ for all effects $E(X){\in}\EE$, (ii) $q(U){=}1$, and (iii) $q({\sum_i}E(X_i)){=}{\sum_i} q(E(X_i))$ for all sequences of effects in $\EE$ that satisfy ${\sum_i}E(X_i){\leqslant} U$, must be of the form $q(A){=}\inprod{A}{B}$ for all $A{\in}\V$, for a unique $B{\in}\V$ which is normalized in the sense that $\inprod{U}{B}{=}1$.
\end{theorem}
As presented in Appendix~\ref{app:gleason}, Theorem~\ref{th:Gleason} is simply proven by extending probability measures on $\EE$ to the whole $\V$ and using Riesz's representation theorem.

Given effect space $\EE$, the state space of the GPT can be delineated using Theorem~\ref{th:Gleason} in conjunction with the {\it no-restriction hypothesis}~\cite{Chiribella2010}, i.e. that given PVVM space $\EE$ {\it all} definable probability measures ($q$'s) on it correspond to physically valid states, as
\begin{equation}\label{eq:S}
\St:=\{\varrho\in\V{|}\inprod{E(X)}{\varrho}\geqslant 0~\forall E(X)\in\EE{,}\inprod{U}{\varrho}=1\}.
\end{equation}
\noindent We denote a GPT by its pair of PVVM and state space as $\T{:=}(\EE{,}\St)$.

Here, instead of {\it a priori} assuming the probability rule of GPTs, we have derived it from a set of reasonable assumptions following in footsteps of Gleason~\cite{Gleason1957}, Busch~\citep{Busch2003}, and Caves {\it et al}~\cite{Caves2004}.
Moreover, it is argued, for example in Refs.~\cite{Barnum2010,Chiribella2011,Janotta2013}, that the no-restriction hypothesis is of no physical basis and thus, it is desirable to drop it from GPT constructions.
We note that, such a ``relaxation" comes at a price: one has to {\it assume a priori} also the state assignments, meaning trading one assumption for another.
Indeed, there is no objection to not imposing the no-restriction hypothesis, however, we have shown in Appendix~\ref{app:subGPT} that any GPT that does not satisfy the no-restriction hypothesis is obtained as a {\it subtheory} of possibly (infinitely) many GPTs that do satisfy it by imposing appropriate further constraints, thus named a subGPT.
To give an example, consider quantum mechanics as a specific GPT that satisfies the no-restriction hypothesis and from which the subGPT of Gaussian quantum mechanics is obtained by restricting the effect and state spaces to those elements possessing Gaussian Wigner representations~\citep{Bartlett2012}.
Notably, quantum theory is {\it not} the unique GPT containing Gaussian quantum mechanics, for it is also a subtheory of classical statistical (or Liouville) mechanics~\cite{Bartlett2012}.

Within the rest of this paper, we use GPT and subGTP to distinguish between theories that do and do not comply with the no-restriction hypothesis, respectively.


\begin{figure}[t!]
\begin{center}
  \includegraphics[width=1\columnwidth]{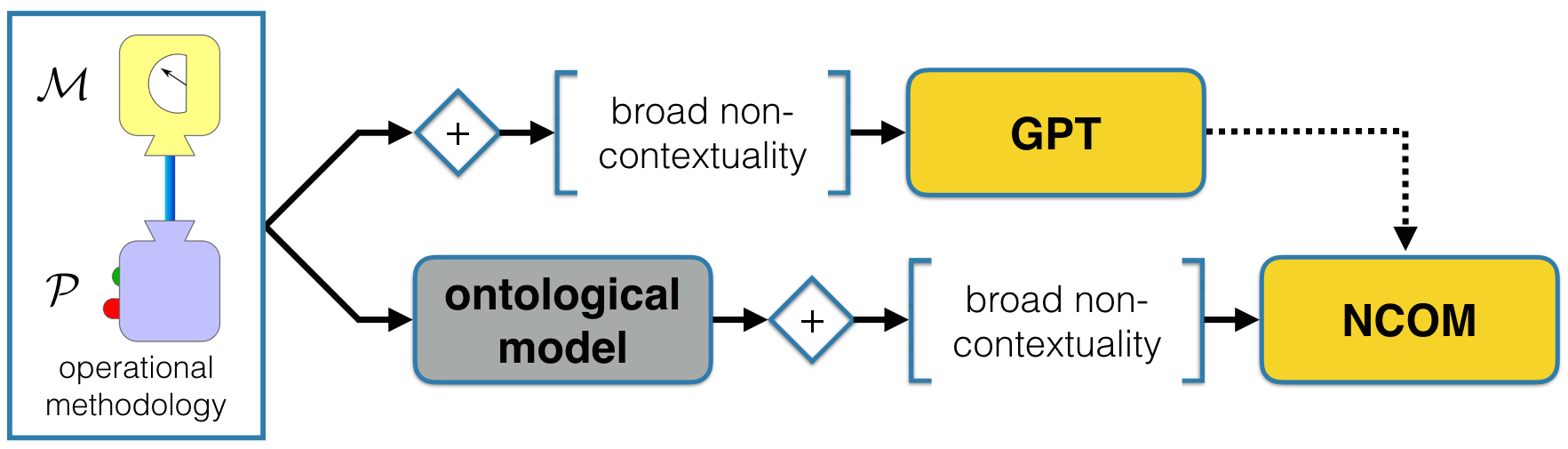}
\end{center}
  \caption{The diagrammatic representation of different model constructions for operational descriptions of experiments.
  The dotted arrow represents the detour approach for building NCOMs for GPTs.
  A GPT that admits an NCOM is called ontologically noncontextual.
  }
  \label{fig:Diagram}
\end{figure}

\section{Broad (Non)Contextuality}
Irrespective of which approach is adopted to explain experimental data, be it an ontological model or a (sub)GPT, {\it broad noncontextuality} (as defined below) is a desirable hypothesis about the description.
Beginning with the {\it statistical equivalence} assumption~\cite{Spekkens2005,HolevoBook}, two preparations $\pp_1{,}\pp_2{\in}\PP$ are statistically indiscernible and equivalent, $\pp_1{\cong}\pp_2$, if and only if for every measurement procedure $\MM{\in}\M$ and every event $X{\in}\omega$ it holds that $p(X{|}\pp_1{,}\MM){=}p(X{|}\pp_2{,}\MM)$.
Similarly, two measurements $\MM_1{,}\MM_2{\in}\M$ are statistically indiscernible and equivalent, $\MM_1{\cong}\MM_2$, if and only if for every preparation procedure $\pp{\in}\PP$ and every event $X{\in}\omega$ it holds that $p(X{|}\pp{,}\MM_1){=}p(X{|}\pp{,}\MM_2)$.
These relations partition the collections of preparations and measurements into equivalence classes $e(\pp)$ and $e(\MM)$ for each preparation $\pp$ and measurement $\MM$.
The particular way in which a state or measurement is experimentally realized corresponds to an element within an equivalence class and it is called a {\it context}.
The {\it broad noncontextuality hypothesis} states that our models of physical phenomena aiming {\it only} at reproducing the statistics should depend only on equivalence classes rather than individual contexts, because statistics do not carry any information about the latter; see e.g. Ref.~\cite{Caves2004}.

An ontological model which is noncontextual in the broad sense is called a {\it noncontextual ontological model (NCOM)} and satisfies
\begin{equation}\label{eq:NCOM}
\begin{split}
    &\pp_1\cong\pp_2 \Leftrightarrow \mu_{\pp_1}=\mu_{\pp_2},\\
    &\MM_1\cong\MM_2 \Leftrightarrow \{\xi_{\MM_1}(X|\lambda)\}=\{\xi_{\MM_2}(X|\lambda)\}.
\end{split}
\end{equation}

Similarly, broad noncontextulaity of (sub)GPTs reads as
\begin{equation}
\begin{split}
    &\pp_1\cong\pp_2 \Leftrightarrow \pp_1,\pp_2\mapsto \varrho,\\
    &\MM_1\cong\MM_2 \Leftrightarrow \MM_1,\MM_2 \mapsto \{E(X)\}.
\end{split}
\end{equation}
Note that, broad noncontextuality is built-in to our GPT and subGPT constructions.
In general, however, it is possible to construct (sub)GPTs that do not respect the broad noncontextuality hypothesis by allowing for context-dependent effect and state assignments.
Importantly, it follows from our constructions here that every NCOM is a (sub)GPT while the converse is not true.
In this Letter, our aim is to determine which (sub)GPTs do and which ones do not admit an NCOM; see Fig~\ref{fig:Diagram}.

\section{Ontological (Non)Contextuality of (Sub)GPTs}\label{subsec:GPTcontextuality}

As we have inferred from experiments in the quantum regime, that to take a {\it direct} route and build an NCOM to describe all possible physical experiments seems very unlikely.
Yet we may ask if it is possible to take a detour, as shown in Fig~\ref{fig:Diagram}, and construct ontological models of (sub)GPTs, noting that, with appropriate care, such models will inherit the noncontextuality from the theory leading to NCOMs.
For quantum theory as a GPT the answer to our question is in negative~\cite{Spekkens2008,Ferrie2009}. 
In case of a generic (sub)GPT, we replace the preparation and measurement procedures $\pp$ and $\MM$ in Eqs.~\eqref{eq:onticstate},~\eqref{eq:onticmeasure}, and~\eqref{eq:onticBayes}, with their representatives in the theory, $\varrho{\in}\St$ and $\{E(X)\}{\subset}\EE$, respectively.
Hence, there should exist {\it injective} maps $\eta{:}\St {\to} \Y$ and $\zeta{:}\EE {\to} \Q$ that assign the {\it unique} ontic state $\eta_{\varrho}$ and ontic measurement $\zeta_{E}$ to each state vector $\varrho$ and PVVM $E$, respectively, such that for all $\lambda {\in} \Upsilon$ and all events $X{\in}\omega$,
\begin{equation}\label{eq:TOntPos}
\eta_{\varrho}(\lambda)\geqslant 0, \quad \zeta_{E}(X{|}\lambda)\in [0{,}1],
\end{equation}
and satisfy
\begin{equation}\label{eq:TOntNorm}
\dmu{\Upsilon}{\lambda\eta_{\varrho}(\lambda)} =1, \quad \text{and}\quad \forall \lambda \quad \zeta_{E}(\Omega{|}\lambda)=1.
\end{equation}
The probability of a particular event $X$ in a measurement $\MM$ given the preparation $\pp$ should then be obtained as
\begin{equation}\label{eq:TOntProbRule}
p(X{|}\pp{,}\MM) {=} p(X{|}\varrho,E) {=} {\dmu{\Upsilon}{\lambda\eta_{\varrho}(\lambda)}} \zeta_{E}(X{|}\lambda).
\end{equation}

Within recent literature a \textit{noncontextual (sub)theory} usually refers to \textit{a (sub)theory that admits an NCOM}~\cite{Spekkens2005}.
For clarity, here we call such a (sub)theory {\it ontologically noncontextual} and reserve the term {\it noncontextuality} for the broader notion.
We note also that any ontologically noncontextual (sub)theory is noncontextual while the converse is not true.
For instance, quantum theory satisfies broad noncontextuality hypothesis but, it is not ontologically noncontextual as it contains effects and states that do not possess a unique convex decomposition in terms of the respective extremal elements~\cite{Spekkens2008,Ferrie2009}.
The latter is clarified by the following analysis.

\subsection{GPTs}\label{subsec:GPTbothS&NRH}

We now analyze ontological noncontextuality of GPTs.
Recall that the maps $\mu$ and $\xi$, and hence $\eta$ and $\zeta$, are convex linear.
Using the fact that $\St$ and $\EE$ both span $\V$, $\eta$ and $\zeta$ can uniquely be extended to the whole space $\V$.
Then, using Riesz's theorem we find that they must be of the forms $\eta_{\varrho}(\lambda){=}\inprod{\varrho}{F(\lambda)}$ and $\quad \zeta_{E}(X{|}\lambda){=}\inprod{E(X)}{D(\lambda)}$ for $F(\lambda),D(\lambda){\in}\V$.
Satisfying Eq.~\eqref{eq:TOntNorm} then requires that $\dmu{\Upsilon}{\lambda F(\lambda)} {=} U$ and $\inprod{U}{D(\lambda)}{=}1$ for all $\lambda\in\Upsilon$.
Thus, $\F{:=}\{F(\lambda)\}$ resembles a PVVM whereas $\D{:=}\{D(\lambda)\}$ is a subset of GPT's state space.
The probability rule of Eq.~\eqref{eq:TOntProbRule} then implies that $\F$ and $\D$ are dual frames~\cite{Ferrie2008,Ferrie2009} for representation of vectors in $\V$ so that
\begin{equation}\label{eq:FDFexpSE}
\varrho {=}{\dmu{\Upsilon}{\lambda\eta_{\varrho}(\lambda)}}D(\lambda),\text{and}~
E(X) {=} {\dmu{\Upsilon}{\lambda F(\lambda)}}\zeta_{E}(X{|}\lambda).
\end{equation}
After taking into account that state vectors are dual to measurement vectors and a few more simple steps (see Appendix~\ref{app:GPTNCOM}) we find that $\F$ and $\D$ must be generating sets of closed convex sets $\EE$ and $\St$, respectively, i.e., $\EE{=}\overline{{\rm conv}\F}$ and $\St{=}\overline{{\rm conv}\D}$.
In light of Eq.~\eqref{eq:FDFexpSE}, we conclude from the latter that the extremal or {\it pure} states $\varrho{\in}\D$ and the extremal or {\it sharp} effects $E(X){\in}\F$, i.e. states and effects that cannot convexly be decomposed into other states and effects, must be represented by Dirac delta measures over the ontic space $\Upsilon$, that is, (i) $\D{\ni}\varrho{\xmapsto{\eta}}\delta_{\lambda_{\varrho}}(\lambda)$ for some $\lambda_{\varrho}{\in}\Upsilon$ and (ii) $\F{\ni} E(X){\xmapsto{\zeta}}\delta_{\lambda_{E(X)}}(\lambda)$ for some $\lambda_{E(X)}{\in}\Upsilon$, where $\delta_a(\beta)$ equals $0$ if $a{\notin}\beta$ and equals $1$ if $a{\in}\beta$ for any measurable subset $\beta$.

The condition (i) is referred to as {\it ontic determinism}, which means a {\it pure} preparation~\cite{Chiribella2010} represented by a pure state determines the ontic value $\lambda$ completely.
Accordingly, we call condition (ii) {\it outcome determinability}, that is, in {\it sharp} measurements~\cite{Chiribella2014} represented by sharp effects of the theory, specifying the ontic variable determines the outcome of the measurement with certainty.
Ontic determinism and outcome determinability enforce that in any ontologically noncontextual GPT state and effect vectors possess {\it unique} decompositions into nonrefinable extremal elements.
Specifically, it is clear from Eq.~\eqref{eq:FDFexpSE} that requiring a unique ontic representation $\eta_\varrho (\lambda)$ for a state vector $\varrho$ simply means a unique convex decomposition into pure states.
Considering sharp effects that cannot be refined (or {\it atomic} effects~\cite{Chiribella2010}) and denoting their collection by $\EE\s{nr}{\subset}\F$, it follows also that the convex decomposition of any effect into elements of $\EE\s{nr}$ must be unique.
The following theorem characterizes GPTs for which these criteria are met and thus, they are ontologically noncontextual.
For a detailed proof, please see Appendix~\ref{app:GPTNCOM}.

\begin{theorem}\label{th:GPTNCOM}
A GPT is ontologically noncontextual if and only if its pure states and nonrefinable sharp effects each form a {\it complete} basis for the space $\V$.
Equivalently, the GPT must be simplicial meaning that $\St$ and $\overline{{\rm conv}\EE\s{nr}}$ are simplexes.
\end{theorem}

\subsection{SubGPTs}\label{subsubsec:GPTSpanNoNRH}

As we discussed earlier, subGPT are GPTs with extra restrictions on their effect and state spaces.
We now determine the condition for a subGPT to admit an NCOM.
\begin{theorem}\label{th:GPTnecsuf}
Any subGPT $\T\s{sub} {=}(\EE\s{sub},\St\s{sub})$ over $\V\s{sub}$ admits an NCOM if and only if it can be thought of as a subtheory of an ontologically noncontextual GPT $\T {=}(\EE,\St)$ over $\V$ and $\dim\V\s{sub}{=}\dim\V{=}\card\Upsilon$.
\end{theorem}
The proof of Theorem~\ref{th:GPTnecsuf} as detailed in Appendix~\ref{app:GPTnecsuf}, similar to that of Theorem~\ref{th:GPTNCOM}, uses the extensions of maps $\eta$ and $\zeta$ to the whole $\V\s{sub}$ which partially explains the condition $\dim\V\s{sub}{=}\dim\V$.
Further clarification is also provided via two examples in Appendix~\ref{app:Examples}.
On one hand, we have shown that Spekkens' toy model~\cite{Spekkens2007} is a $4$-dimensional subGPT admitting an NCOM in four dimensions hence considered to be a classical subGPT; cf. Appendix~\ref{subapp:SpekkensToy}.
On the other hand, the stabilizer rebit theory is shown to be a $3$-dimensional subGPT the smallest ontological model admitted by which is $4$-dimensional and thus, ontologically contextual; cf. Appendix~\ref{subapp:rebit}.

From our discussion, it follows that the stabilizer rebit subGPT cannot be regarded classical despite that it is classically simulatable in $4$-dimensions.
The caveat of the $4$-dimensional ontological model is that the excess dimension gives room for multiple ontic states that are statistically indiscernible under the subGPT's measurements, hence violating the broad noncontextuality hypothesis of Eq.~\eqref{eq:NCOM}, specifically, $\pp_1{\cong}\pp_2 {\nRightarrow} \mu_{\pp_1}{=}\mu_{\pp_2}$; see Appendix~\ref{subapp:rebit}.
Intuitively, this is the case whenever $\dim\V{>}\dim\V\s{sub}$.
Classicality of (sub)GPTs thus requires both their classical simulatability and that there are no in-principle inaccessible parameters at the ontological level, i.e. the (sub)GPT is ontologically complete.
The latter is in accordance with Leibniz's methodological principle of the ontological identity of empirical indiscernibles~\cite{Spekkens2019,Mazurek2016}.
We emphasize that this constraint does {\it not} conflict with NCOM approaches to experimental data presented e.g. in Refs.~\cite{Schmid2018} and~\cite{Schmid2018PRA}; see Appendix~\ref{app:conflict} for a further discussion.

%

\section{Resources and Contextuality in GPTs}
We now consider the concept of resources at a fundamental level.
Imagine experiments with sets of preparations $\PP$ and measurements $\M$, and an experimenter who has devised a GPT $\T{=}(\EE{,}\St)$ capable of explaining the statistics obtained in her experiments.
We call the set of all possible measurements $\EE$ {\it free measurements} and the set of all possible preparations $\St$ {\it free states}.
We remark that in quantum resource theories~\cite{Chitambar2018,Horodecki2009,Brandao2013,Horodecki2013,Brandao2015,Faist2015,Lostaglio2015,Gour2015,Aberg2006,
Baumgratz2014,Winter2016,Streltsov2017,Gour2008,Gour2009,Piani2016,Rivas2014,Rivas2015} free states and measurements are subsets of all possible states and measurements.
Here, in contrast, all possible preparations and measurements are defined to be free and non-free (or {\it resourceful}) ones are those yet to be discovered and thus not specified in the GPT.

Now, suppose that the experimenter discovers another preparation or measurement procedure that was not previously known to exist.
Naturally, she has to come up with a new theory $\T^\star{:=}(\EE^\star{,}\St^\star)$ to reproduce also the new measurement data.
We ask how $\T^\star$ compares to $\T$ in terms of (non)classicality.
We only consider the case wherein the extended theory $\T^\star$ is required to satisfy the no-restriction hypothesis.
Then, by Theorem~\ref{th:GPTNCOM}, given the new sets of measurements and preparations, $\M^\star$ and $\PP^\star$, if the nonrefinable sharp measurements of the new GPT form a complete set of PVVMs over a vector space $\V^\star$, then the GPT will be ontologically noncontextual.
In such scenarios, even though the newly discovered element is a resource, all the measured statistics can be explained in classical terms.
Hence, we call such a bonus procedure a {\it classical} resource.
On the other hand, a resourceful element may enforce a nonclassical extension of the older theory, where it is called a {\it nonclassical} resource.
The conditions for a single resource to dictate the use of a nonclassical model for an explanation of possible statistics is provided below and proved in Appendix~\ref{app:GPTRes}. 

\begin{theorem}\label{th:GPTRes}
Suppose that the set of free measurements $\M$ and preparations $\PP$ are represented by PVVMs $\EE$ and states $\St$ in some classical GPT $\T{=}(\EE{,}\St)$.
Given a single nonrefinable bonus measurement $\MM^\star$ (preparation $\pp^\star$), the followings are equivalent: 
(i) $\T^\star$  is ontologically contextual;
(ii) $\MM^\star$ ($\pp^\star$) is a nonclassical resourceful measurement (preparation);
(iii) The PVVM $E^\star$ (state $\varrho^\star$) nonconvexly overcompletes the nonrefinable effects $\EE\s{nr}$ (states $\St$) into $\EE\s{nr}^\star$ ($\St^\star$);
(iv) $E^\star$ ($\varrho^\star$) lies within $\V^\star{=}\V$ but $E^\star{\notin}\EE$ ($\varrho^\star{\notin}\St$).
\end{theorem}


\section{Conclusions}
We considered the phenomenon of generalized contextuality in general probabilistic theories (GPTs) and showed that any GPT satisfying the no-restriction hypothesis is ontologically noncontextual if and only if it is simplicial.
We also discussed extensively the case of subGPTs that do not comply with the norestriction hypothesis.
Our results shows that any GPT can at most subsume two of the three properties of satisfying the no-restriction hypothesis, ontological noncontexuality, and possessing multiple nonrefinable measurements.
Some examples to each possibility already exist, e.g. dropping the no-restriction hypothesis that results in subGPTs as Gaussian quantum mechanics~\cite{Bartlett2012}, giving up on the ontological noncontextuality as in full quantum theory, or capitulating incompatibility of measurements as in classical mechanics.

A secondary aspect of our work is to provide a new route towards the characterization of the nonclassical power of individual operational elements in information processing protocols by noting that many nonclassical advantages, though describable by quantum formalism, do not depend on the specifically assumed underlying theory.
As an example, in quantum computations we can draw conclusions about their nonclassicality (similarly to the Bell scenario) by merely relying on the classical inputs and outputs and the hardness promises of the computational complexity theory.
We thus arrive at two conclusions.
First, a nonclassical advantage in a quantum scenario also implies a nonclassical advantage in any postquantum theory including those formulated within the GPT framework.
Second, the resource formalism suitable for explaining the nonclassical advantages should describe a fundamental theory-independent property, i.e. one that can be certified merely by relying on the measurement statistics.
Contextuality is one such a property that ipso facto forbids a classical description of the processes. 
Combining these two, in our opinion, the study of contextuality of GPTs and the contextual power of single operational elements is a good candidate for a new approach towards a resource theoretic resolution to the fundamental problem of sufficient resources for quantum computations.


\begin{acknowledgments}
The author is thankful to John Selby and Rob Spekkens for fruitful discussions.
We acknowledge comments by Michael Hall and Markus M\"{u}ller on an earlier draft of this paper.
We also acknowledge support and resources provided by the Royal Commission for the Exhibition of 1851, AQTION project (820495) funded by the European Union Quantum Technology Flagship, and the S\^{e}r SAM Project at Swansea University, an initiative funded through the Welsh Government's S\^{e}r Cymru II Program (European Regional Development Fund).
\end{acknowledgments}

\paragraph*{Note added.---} During the preparation of this paper, we became aware of two independent works by Schmid~{\it et al}~\cite{Schmid2019} and Barnum and Lami~\cite{Barnum}.
In the former, the concept of {\it simplex-embeddable} GPTs, which is equivalent to our notion of ontologically noncontextual subGPTs, is introduced.
Accordingly, a result similar to Theorem~\ref{th:GPTnecsuf} is presented.
Here, we further developed the idea by identifying the relationship between the dimensionalities of subGPTs and ontological models.
The latter work discusses a result overlapping with Theorem~\ref{th:GPTNCOM}.

\appendix

\section{A brief review of measures} \label{app:revmeas}

The way we lean about the notion of ``size'' in school is a very basic one with value assignments that are real or complex numbers.
For example, we learn to calculate the area (or the two-dimensional volume) of a rectangle by multiplying its sides.
We are thought later on that the volume under some continuous function can be evaluated by Riemannian integration.
This very {\it sensible} approach is methodological rather than conceptual.

In order to abstractize and henceforth generalize the notion of ``size'', we need to pursue the usual mathematical procedure of creating a {\it mathematical structure}.
This route consists of defining some proper sets of objects, some minimal axioms on those sets that shape the structure, and potentially some suitable functions on those sets.
Once this is done to capture ``size'', we get to a mathematical structure known as {\it measure theory}.
In the following, we briefly review this procedure for interested physicists.

Suppose a set $\Omega$ is given.
It turns out that ``sizes'' cannot be defined for one such arbitrary set.
Therefore, instead of defining ``sizes'' for elements of $\Omega$, we will define them for elements of another set generated by $\Omega$, called the $\sigma$-algebra over $\Omega$.

\begin{definition}
Let $\Omega$ be a nonempty set.
A $\sigma$-algebra for $\Omega$ is a subset $\omega$ of the power set $2^\Omega$ (i.e. the set of all subsets of $\Omega$) that satisfies the following axioms:
\begin{enumerate}[(i)]
\item $\omega$ contains $\Omega$;
\item If $X\in\omega$ then $\Omega\backslash X \in \omega$;
\item For any countable sequence $\{X_i\in omega\}_{i\in\mathbb{N}}$ it holds that $\cup_{i}X_i\in \omega$.
\end{enumerate}
\end{definition}

\begin{definition}
The pair $(\Omega,\omega)$ of an underlying set together with a $\sigma$-algebra on it is called a {\it measurable} space.
\end{definition}
\noindent Now, one might be able to develop a sense of why the axioms of the $\sigma$-algebra make it possible to meaningfully assign ``sizes'' to its elements.
The main idea is that we do have, in a sense, the notions of {\it complementarity} (axiom (ii)) and {\it union} (axiom (iii)) of elements.
The first one allows us to understand the {\it relative} ``sizes'' of different elements, including the relative ``size'' to the whole collection of elements (axiom (i)).
The second one allows us to grasp the meaning of the ``size'' of a joint collection of elements.
We can now precisely state what we mean by ``size''.

\begin{definition}
Given a measurable space $(\Omega,\omega)$, a {\it measure} $\mu$ is a function $\mu:\omega\rightarrow[0,\infty]$ such that
\begin{enumerate}[(i)]
\item $\mu(\emptyset)=0$;
\item For any sequence of pairwise disjoint elements $\{X_i\in\omega\}_{i\in\mathbb{N}}$ it holds that
\begin{equation}
\mu(\cup_i X_i)=\sum_i \mu(X_i).
\end{equation}
\end{enumerate}
\end{definition}
\noindent Here, by pairwise disjoint elements we mean $X_i\cap X_j=\emptyset$ whenever $i\neq j$.
Again, a measure has a simple interpretation. 
For convenience, we need the empty set to have a ``size'' of zero.
This way, we map the identity element of union of sets to the additive identity element of $\R$.
This is done by axiom (i).
Next, we complete the similarity between set union and addition over $\R$ by condition (ii), which is called the {\it additivity} property.

It must be now clear that $\mu$ together with its underlying space deliver all the properties we expect from ``size'' assignments to objects in a meaningfully general sense.

\begin{definition}
The triplet $(\Omega,\omega,\mu)$ is called a {\it measure (or measured)} space.
\end{definition}

\begin{definition}
The elements $X$ of the $\sigma$-algebra $\omega$ are called {\it measurable subsets} of $\Omega$.
\end{definition}

A very useful measure which is also used within the main text is Dirac's delta measure.
Given the measurable space $(\Omega,\omega)$, Dirac delta measure is defined as
\begin{equation}
\delta_X(x)=
\left\{\begin{matrix}
0 & x\notin X \\ 
1 & x\in X,
\end{matrix}\right.
\end{equation} 
where $X\in \omega$ is a measurable subset of $\Omega$ and $x\in\Omega$.
Dirac delta measure has the ``familiar'' property that given a measure $f$ on $(\Omega,\omega)$ we have
\begin{equation}
\int_\Omega f(x)\delta_X(x) dx = f(X).
\end{equation} 

Consider now the case where $\Omega$ is the set of all outcomes of some measurement on some physical system.
We can then define {\it events} that are subsets of all possible outcomes.
For instance, in flipping a coin the outcomes are `heads' or `tails' and the events are $\{\emptyset$,`heads',`tails',`heads $\lor$ tails'$\}$. 
The event $\emptyset$ is impossible whereas the event `heads $\lor$ tails' is certain.
It is straightforward to see that the event set can be identified with the power set of the set of measurement outcomes, hence it can be regarded as a $\sigma$-algebra.
As a result, each event is a {\it measurable subset} of the set of all outcomes.
This is, in fact, the case for all measurements on all physical systems, motivating us to define the following measure.
\begin{definition}
Given a measurable space $(\Omega,\omega)$, a {\it probability measure} $p$ is a measure whose range is the closed interval $[0,1]$ and satisfies $p(\Omega)=1$.
\end{definition}
\noindent Clearly, $p$ carries the interpretation of a probability distribution over the space of events assigning a ``size'' to each event that is equivalent to our belief of occurrence of that event.

As mentioned in the main text, indeed, there is no particular reason to restrict the range of measure functions to that of real intervals.
Once we replace the range interval $[0,1]$ of a probability measure with a subset of a vector space we obtain probability measure that its values are vectors
rather than real numbers, that is a {\it probability vector-valued measure} (PVVM) as defined in the main text.
Note that the latter holds provided we can identify a fixed identity element ($U$) within the vector space.

\section{Proof of Theorem~\ref{th:Gleason}} \label{app:gleason}

Following the approach of Refs.~\cite{Busch2003,Caves2004}, from the additivity property (iii) and that given any $E\in\EE$, the effect $pE$ for any real number $p\in[0,1]$ also belongs to $\EE$, it follows that the map $q$ is homogeneous over nonnegative rational numbers, i.e., $q(mE/n)=mq(E)/n$ for any $E\in\EE$ and $m,n\in\Z^+$.

Next, suppose that $E\in\EE$ and $\alpha,\beta\in [0,1]$ with $\alpha<\beta$.
Thus, $\alpha E$ and $\beta E$ belong to $\EE$ and $\alpha E<\beta E$.
It is also clear that $E':=\beta E-\alpha E=(\beta-\alpha)E$ belongs to $\EE$.
Since $\beta E=E'+\alpha E$ implies $q(\beta E)=q(E')+q(\alpha E)$ and $q(E')\geqslant 0$ by requirement (i), we have $q(\alpha E)\leqslant q(\beta E)$.
That is, the map $q$ preserves the order of elements within $\EE$.

Now consider a pair of increasing and decreasing sequences of rational numbers in the $[0,1]$ interval, $(\alpha_i)_i$ and $(\beta_i)_i$, respectively, where both converge to the same irrational value $\gamma\in[0,1]$.
From order preserving property of $q$ and its homogeneity over rational numbers we obtain $\alpha_i q(E)=q(\alpha_iE)\leqslant q(\gamma E)\leqslant q(\beta_i E)=\beta_i q(E)$. 
Then by the pinching theorem we have $q(\gamma E)=\gamma q(E)$, that is, the map $q$ must be linear.

Since $q$ is a convex-linear functional that is defined on a spanning convex subset ($\EE$) of a vector space ($\V$), it can be uniquely extended to a linear functional on the whole vector space ($\V$).
Finally, by Riesz's theorem, this linear functional can be written as the inner product $q(A)=\inprod{A}{B}$ for a unique vector $B\in\V$.

The normalization of $B$ follows simply from requirement (ii) as $q(U)=\inprod{U}{B}=1$.
\qed

\section{GPTs as containers for subGPTs} \label{app:subGPT}

\begin{prop}\label{prop:NRH}
Any GPT $\T\s{sub}$ defined on a vector space $\V$ and identified by the pair $\T\s{sub}{:=}(\EE\s{sub}{,}\St\s{sub})$ of its physically allowed PVVMs and states that does not satisfy the no-restriction hypothesis is a {\it subtheory} of possibly (infinitely) many extended GPTs that do satisfy it.
\end{prop}

\begin{figure}[t!]
\begin{center}
  \includegraphics[width=0.8\columnwidth]{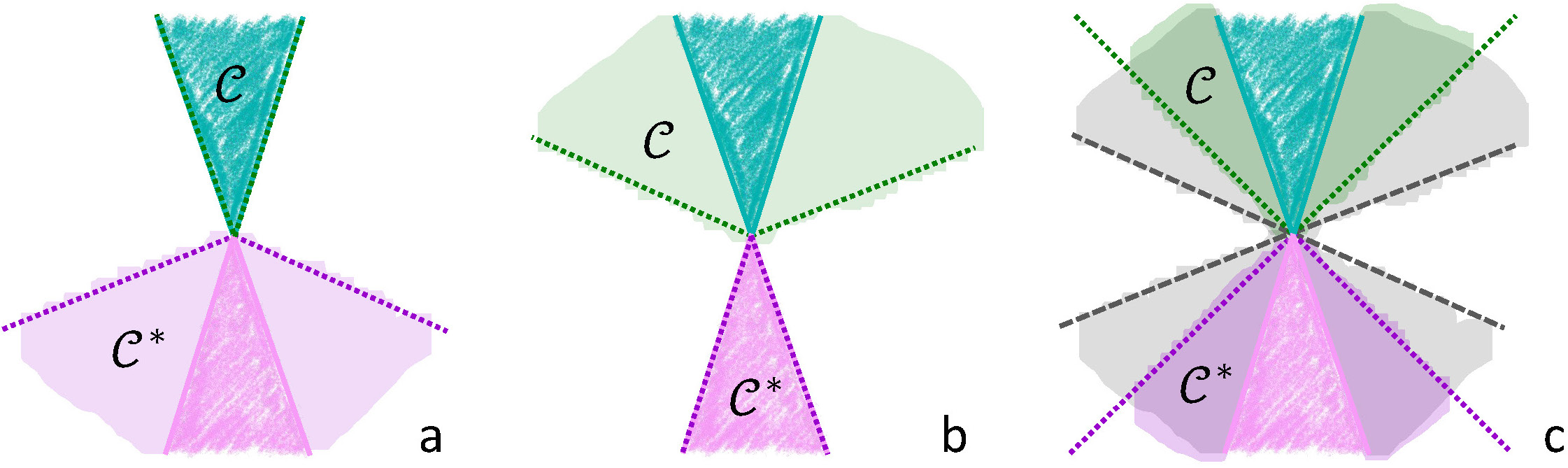}
\end{center}
  \caption{A heuristic representation of the cones containing three possible GPTs sharing a subGPT.
  As described in the proof of Proposition~\ref{prop:NRH},
  (a) $\T=(\EE\s{sub},\St)$ where $\St\s{sub}\subseteq\St$;
  (b) $\T'=(\EE,\St\s{sub})$ where $\EE\s{sub}\subseteq\EE$;
  (c) $\T''=(\EE'',\St'')$ where $\EE\s{sub}\subseteq\EE''\subseteq\EE$ and $\St\s{sub}\subseteq\St''\subseteq\St$.
  While the subGPT does not satisfy the no-restriction hypothesis, the extended GPTs do satisfy it, i.e. their effect and state spaces are dual to each other.
  Note that the effect or the state space of all three GPTs can be restricted in such a way to reproduce the given subGPT.
  }
  \label{fig:GPTIncl}
\end{figure}

\paragraph*{Proof.}
It follows from violation of the no-restriction hypothesis that $\St\s{sub}\subset\St$, where $\St$ is defined as 
\begin{equation*}
\St:=\{\varrho\in\V{|}\inprod{\varrho}{E(X)}\geqslant 0~\forall E(X)\in\EE\s{sub}{,}\inprod{U}{\varrho}=1\}.
\end{equation*}
Hence, $\T\s{sub}$ is a subtheory of $\T:=(\EE\s{sub},\St)$, where $\T$ satisfies the no-restriction hypothesis; see Fig.~\ref{fig:GPTIncl}a.
Alternatively, one can fix the state space and define the set of effects as
\begin{equation*}
\EE:=\{E\in\V{|}\inprod{\varrho}{E}\in[0,1]~\forall \varrho\in\St\s{sub}\}.
\end{equation*}
Clearly this time $\EE\s{sub}\subset\EE$ and thus $\T\s{sub}$ is a subtheory of $\T':=(\EE,\St\s{sub})$, where $\T'$ satisfies the no-restriction hypothesis; see Fig.~\ref{fig:GPTIncl}b.

Finally, any GPT $\T'':=(\EE'',\St'')$ such that $\EE\s{sub}\subset\EE''\subseteq\EE$, $\St\s{sub}\subset\St''\subseteq\St$, and $\EE''$ and $\St''$ are dual sets, contains $\T\s{sub}$ as its subtheory; see Fig.~\ref{fig:GPTIncl}c.
\qed

\section{Generators of GPT state and effect spaces} \label{app:GPTgenerators}

\begin{lemma}\label{lem:DualGen}
The pair $(\F{,}\D)$ mapping a GPT that satisfies the no-restriction hypothesis to an NCOM must be the generating sets of the pair of closed convex sets $(\EE{,}\St)$. That is, $\EE{=}\overline{{\rm conv}\F}$ and $\St{=}\overline{{\rm conv}\D}$.
\end{lemma}

\paragraph*{Proof.}
In order to have an NCOM, it is first required that $\zeta_{E}(X|\lambda)=\inprod{E(X)}{D(\lambda)}\geqslant 0$ for all $E\in\EE$ and all $\lambda\in\Upsilon$.
Therefore, in view of the no-restriction hypothesis as in the definition of the GPT's state space in Eq.~(4) of the main text and that $\inprod{U}{D(\lambda)}=1$ for all $\lambda\in\Upsilon$, we find that $D(\lambda)\in \St$, which implies $\overline{{\rm conv}\D}\subseteq\St$.
Second, by imposing the second requirement of a noncontextual ontological model on Eq.~\eqref{eq:FDFexpSE}, that is, $\eta_{\varrho}(\lambda)\geqslant 0$ for all $\varrho\in\St$ and all $\lambda\in\Upsilon$, we see that $\St\subseteq\overline{{\rm conv}D(\lambda)}$.
Combining together, it must hold true that $\overline{{\rm conv}\D}=\St$ and thus, the set $\D$ is a generating set of the GPT's state space $\St$.

Similarly, starting from the requirement $\eta_{\varrho}(\lambda)=\inprod{\varrho}{F(\lambda)}\geqslant 0$ for all $\varrho\in\St$ and all $\lambda\in\Upsilon$ for an NCOM and noting that the set of effects is dual to the set of states with $\dmu{\Upsilon}{F(\lambda)} =U$, we infer that $F(\lambda)\in\EE$ and thus, $\overline{{\rm conv}\F}\subseteq\EE$.
Next, using the fact that $\zeta_{E}(X|\lambda)\geqslant 0$ for all $E\in\EE$ and all $\lambda\in\Upsilon$ in Eq.~(10) of the main text, that is,
\begin{equation}
\begin{split}
\varrho {=}{\dmu{\Upsilon}{\eta_{\varrho}(\lambda)}} ~D(\lambda),~\text{and}~
E(X) {=} {\dmu{\Upsilon}{F(\lambda)}} ~\zeta_{E}(X{|}\lambda),
\end{split}
\end{equation}
we find that $\EE\subseteq\overline{{\rm conv}\F}$.
Together, we have $\EE=\overline{{\rm conv}\F}$, meaning that the set $\F$ is a generating set of the GPT's set of allowed effects $\EE$.
\qed

\section{Proof of Theorem~\ref{th:GPTNCOM}} \label{app:GPTNCOM}

To give the proof of the theorem, we first need to state and prove the following geometrically intuitive lemma.
\begin{lemma}\label{lem:NCLD}
Given a nonconvexly overcomplete basis $\A=\{A_i\}$ of vectors for an ordered vector space $\V$, given that elements of $\A$ belong to the positive pointed generating cone $\CC$, there exists an element $C\in\overline{{\rm conv}\A}$ such that its convex decomposition into elements of $\A$ is not unique.
\end{lemma}
\paragraph*{Proof of Lemma~\ref{lem:NCLD}.}
First, the overcompleteness of $\A$ means that there exists a vector $A_J\in\A$ that is linearly dependent on the elements in $\A\setminus A_J$.
Nonconvexly overcompleteness thus means that $A_J$ possesses a nonconvex decomposition $A_J=\sum_{i\neq J}\alpha_iA_i$ in terms of other elements of $\A$ such that at least one of the expansion coefficients $\alpha_i$ is negative.

Our proof of the Lemma is constructive.
We first show that there exists a vector $B\in\overline{{\rm conv}\A}$ such that $B\notin\overline{{\rm conv}\A \setminus A_J}$ and
\begin{equation}\label{eq:BOp}
B=\sum_{i\neq J} \beta_i A_i,\qquad \sum_{i\neq J}\beta_i = 1.
\end{equation}
Considering the vector $A_J$, if $\sum_{i\neq J}\alpha_i >0$ then we can simply set $B=A_J /\sum_{i\neq J}\alpha_i$.
Otherwise, bearing in mind that due to being an element of a positive cone all $\alpha_i$'s cannot simultaneously be negative, we consider a coefficient $0<\alpha^\star\in\{\alpha_i\}_{i\neq J}$ that corresponds to the operator $A^\star\in\{A_i\}_{i\neq J}$.
Define the operator $B(p):=pA_J + (1-p) A^\star = p \sum_{i\neq J}\alpha_i A_i + (1-p) A^\star$.
Then, for the expansion coefficients of $B(p)$ it holds true that $\sum_{i\neq J}\beta_i (p) = p \sum_{i\neq J}\alpha_i + (1-p)$.
By defining $\overline{\alpha}:=|\sum_{i\neq J}\alpha_i|$ we find
\begin{equation}
\overline{p}:=p=
\left\{\begin{matrix*}[l]
\text{any} ~p\in (0,1), & \text{if}~ \overline{\alpha}=0, \\ 
\frac{1}{1+\overline{\alpha}} & \text{otherwise},
\end{matrix*}\right.
\end{equation}
for which $\sum_{i\neq J}\beta_i (\overline{p})=1$.
Evidently, $B(\overline{p})\in\overline{{\rm conv}\A}$ by construction.
However, because there exists at least one $\beta_i(\overline{p})=\overline{p}\alpha_i<0$, it also holds true that $B(\overline{p})\notin\overline{{\rm conv}\A\setminus A_J}$.
We thus can set $B=B(\overline{p})$.

Given the operator $B$ with the properties as in Eq.~\eqref{eq:BOp}, we consider two sets of indices: $\id^-:=\{i\neq J|\beta_i<0\}$ and $\id^+:=\{i\neq J|\beta_i> 0\}$ and define the operator
\begin{equation}\label{eq:COp}
C:= \frac{1}{N}(B + \sum_{i\in \id^-} |\beta_i|A_i) = \frac{1}{N} \sum_{i\in \id^+} \beta_i A_i,
\end{equation}
in which $N=\sum_{i\in \id^+} \beta_i$.
We see that, both sides of Eq.~\eqref{eq:COp} are convex decompositions of $C$ into elements of $\A$, while only the first decomposition contains the operator $A_J$ (implicit in $B$).
\qed

\paragraph*{Proof of Theorem~\ref{th:GPTNCOM}.}
If the points of $\EE\s{nr}$ form a {\it nonconvexly overcomplete} basis for $\V$, then by Lemma~\ref{lem:NCLD} above, there exists a vector (an effect) $C$ within $\EE$ which is not a coarse-grained effect and yet it possesses a non-unique decomposition in terms of $\EE\s{nr}{\subset}\F$.
This means the effect $C$ possesses a non-unique representation in the ontological model that is not due to coarse-graining.
As a result, such a GPT does not admit an NCOM.
A similar argument holds if the extreme points of $\St$ form a nonconvexly overcomplete basis.
\qed 

\section{Proof of Theorem~\ref{th:GPTnecsuf}} \label{app:GPTnecsuf}

If:
All subGPTs $\T\s{sub}$ of a hypothetical ontologically noncontextual GPT $\T {=}(\EE,\St)$ are ontologically noncontextual provided that $\dim\V\s{sub}{=}\dim\V$. 
To see this, we note that for any subGPT $\T\s{sub}{=}(\EE\s{sub},\St\s{sub})$ it holds true that $\EE\s{sub}\subseteq\EE$ and $\St\s{sub}\subseteq\St$, where $\EE$ and $\St$ are the effect and state spaces of the hypothetical {\it ontologically noncontextual} GPT $\T$.
Then for all elements $E(X)\in\EE\s{sub}$ and $\varrho\in\St\s{sub}$ one can simply use the same ontic assignments $\zeta_E(X|\lambda)$ and $\eta_\varrho(\lambda)$ that one would assign to the states and effects of the hypothetical GPT $\T=(\EE,\St)$.
These assignments trivially reproduce the statistics of the subGPT.
There is, however, a subtle point: the physically allowed measurements are only those defined by the subGPT as $\EE\s{sub}$.
Now, if $\dim\V\s{sub}{<}\dim\V$, then for every state of the subGPT there will exist another hypothetical state of the GPT within $\V\s{sub}^{\perp}\subset\V$ that, if possible, would produce exactly the same statistics under those allowed measurements of the subGPT.
Correspondingly, there will be multiple ontic states that generate exactly the same statistics under the physically possible measurements, hence, $\pp_1{\cong}\pp_2 {\nRightarrow} \mu_{\pp_1}{=}\mu_{\pp_2}$.
In other words, the ontological model will not be satisfying the broad noncontextuality assumption (please see Sec.~\ref{subapp:rebit} below for an explicit example).

The only if direction can be shown as follows.
Assume that the subtheory $\T\s{sub}=(\EE\s{sub},\St\s{sub})$ admits an NCOM.
This means that there exist bijective convex linear maps $\eta\s{sub}$ and $\zeta\s{sub}$ from state and effect spaces of the subtheory to the ontic state and indicator functions over some ontic variable space $\Upsilon$.
We note that we can always assume that the effect and state spaces of the subGPT span its underlying vector space $\V\s{sub}$.
As a result, $\eta\s{sub}$ and $\zeta\s{sub}$ can be uniquely extended to bijective maps $\eta$ and $\zeta$ over the whole $\V\s{sub}$.
We also know that, due to being probability measures on $\Upsilon$, the ontic state and indicator function spaces of the NCOM can be enlarged to ontic state and measurement spaces whose extreme points are Dirac delta measures.
Indeed, the extended ontic spaces are dual to each other and satisfy the no-restriction hypothesis.
In the next step, we simply apply $\eta^{-1}$ and $\zeta^{-1}$, the inverses of the extended maps $\eta$ and $\zeta$, to the extended ontic spaces to obtain extended state and effect spaces $\St\supseteq\St\s{sub}$ and $\EE\supseteq\EE\s{sub}$ in $\V$.
It is immediate that $\dim\V=\dim\V\s{sub}$, and $\EE$ and $\St$ are dual thus the GPT satisfies the no-restriction hypothesis.
Moreover, $\T=(\EE,\St)$ is ontologically noncontextual by construction.
Finally, $\T\s{sub}$ is a subtheory of $\T$.
We emphasize that the GPT $\T$ constructed in this way is only a hypothetical theory in the sense that the physically allowed states and measurements are only those described by the subGPT $\T\s{sub}$.
\qed

\section{Two Examples} \label{app:Examples}

\subsection{Spekkens' toy theory} \label{subapp:SpekkensToy}

As an interesting example of the application of Theorems~\ref{th:GPTNCOM} and~\ref{th:GPTnecsuf}, we analyse the noncontextuality of the Spekkens' toy theory~\cite{Spekkens2007}.

\begin{figure}[t!]
\begin{center}
  \includegraphics[width=0.8\columnwidth]{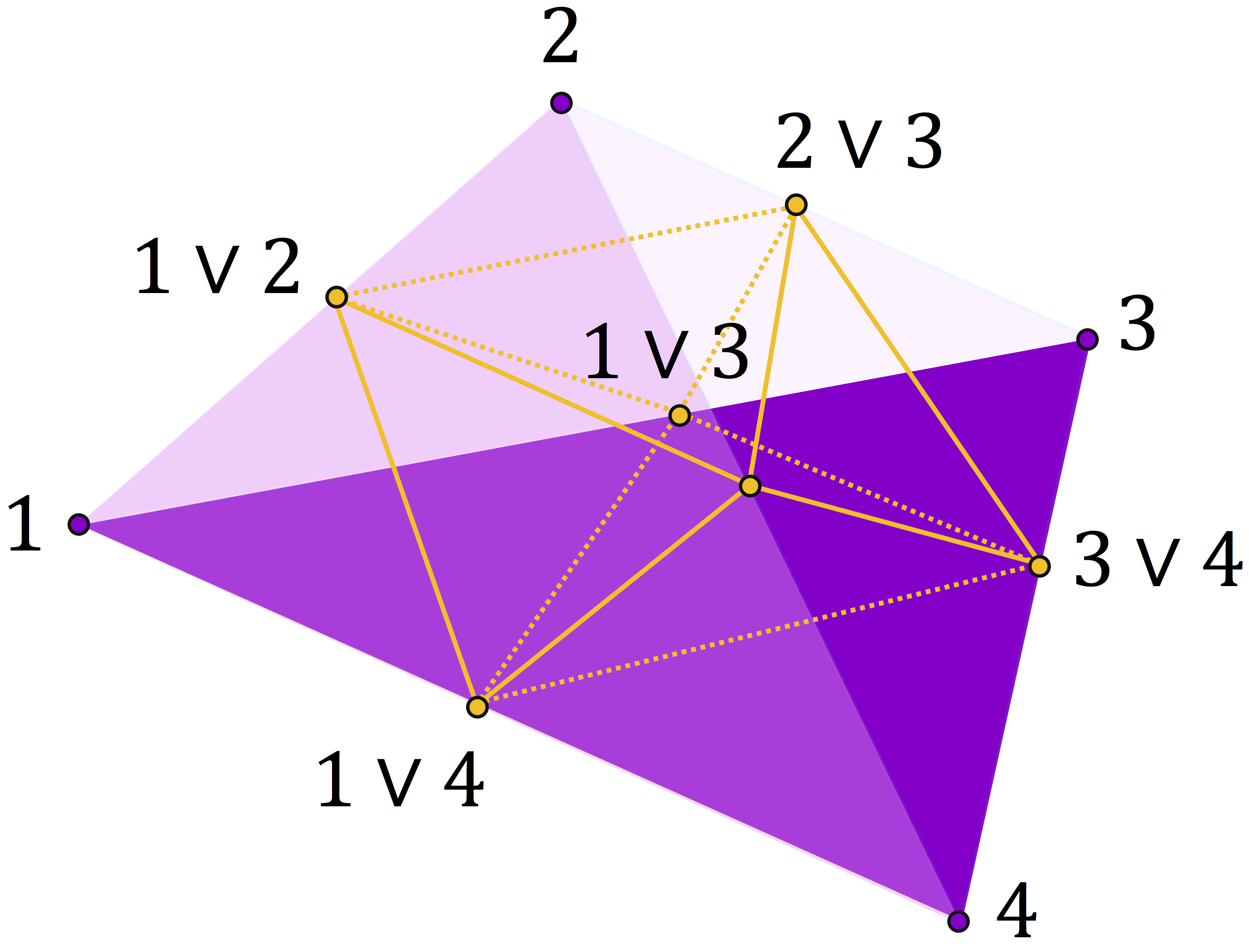}
\end{center}
  \caption{A three-dimensional representation of the (ontic) state and effect spaces of the GPT underlying Spekkens' toy theory.
  This is a simplex whose vertices correspond to vectors spanning $\R^4$.
  The (epistemic) state and effect spaces of the subGPT (i.e. Spekkens' toy theory) is obtained by imposing the knowledge balance principle, which are given by octahedra inside the GPT's state and effect spaces.
  Clearly, while the state and effect spaces of the GPT are dual to each other and satisfy the no-restriction hypothesis, the state and effect spaces of the subGPT are not duals and do not satisfy this hypothesis.
  Nevertheless, Spekkens' toy theory is ontologically noncontextual according to Theorem~\ref{th:GPTnecsuf}, because it is a subtheory of an ontologically noncontextual GPT.
  }
  \label{fig:Spekkens}
\end{figure}

In this model, there exists an elementary system whose pure ontic states are denoted by ``$1$'', ``$2$'', ``$3$'', and ``$4$'' that can be represented by column vectors
\begin{equation}
\eta_1=
\begin{pmatrix}
1\\ 
0\\ 
0\\ 
0
\end{pmatrix},
\eta_2=
\begin{pmatrix}
0\\ 
1\\
0\\ 
0
\end{pmatrix},
\eta_3=
\begin{pmatrix}
0\\ 
0\\ 
1\\ 
0
\end{pmatrix},
\eta_4=
\begin{pmatrix}
0\\
0\\ 
0\\ 
1
\end{pmatrix},
\end{equation}
respectively, i.e. a simplex in $\R^4$; see also Fig.~\ref{fig:Spekkens}.
The measurements on the system can be considered as questions about the state of the system.
Of particular interest are the two outcome measurements that are {\it reproducible}, or {\it repeatable}~\cite{Chiribella2014}, in the sense that if repeated upon a system in a pure state they always output the same outcome.
The set of all such measurements define the sharp effects of the dichotomic measurements, and they are given by ``is the state of the system $1$ or not?'', ``is the state of the system $2$ or not?'', ``is the state of the system $3$ or not?'', and ``is the state of the system $4$ or not?''.
Naturally, these measurements are coarse-grained versions of a single four-outcome nonrefinable measurement that can also be mapped to a simplex in $\R^4$.
This PVVM is given by 
\begin{equation}
\zeta_1=
\begin{pmatrix}
1\\ 
0\\ 
0\\ 
0
\end{pmatrix}\trs,
\zeta_2=
\begin{pmatrix}
0\\ 
1\\
0\\ 
0
\end{pmatrix}\trs,
\zeta_3=
\begin{pmatrix}
0\\ 
0\\ 
1\\ 
0
\end{pmatrix}\trs,
\zeta_4=
\begin{pmatrix}
0\\
0\\ 
0\\ 
1
\end{pmatrix}\trs,
\end{equation}
with the unit element $\zeta_U=(1,1,1,1)$.
Note that the set of all PVVMs which is dual to the state space forms a hypercube. 
As such, according to Theorem~\ref{th:GPTNCOM}, we have a $4$-dimensional classical GPT for the elementary system that satisfies the no-restriction hypothesis.

Given the elementary system and the GPT for it, one can define a {\it canonical measurement set} to be a minimal set of possibly coarse-grained measurements that fully determine the state of the system.
For the elementary system above, one such a set, for example, is given by $\{$``is the state of the system $1$ OR $2$, or not?'', ``is the state of the system $1$ OR $3$, or not?''$\}$.
At this point, it is possible to define the {\it amount of knowledge} to be the maximum number of questions for which the answer is known, varied over all possible canonical measurement sets.

Given a measure of knowledge, an {\it epistemic} restriction can be imposed on the theory, which results in Spekkens' toy theory.
\begin{quotation}
{\it The knowledge balance principle~\cite{Spekkens2007}:} If one has the maximal knowledge, then for every system, at every time, the amount of knowledge one possesses about the ontic state of the system at that time must equal the amount of knowledge one lacks.
\end{quotation}

After enforcing the knowledge balance principle on the original theory, a subtheory will be obtained the pure states of which are given as ``$1 \vee 2$'', ``$1 \vee 3$'', ``$1 \vee 4$'', ``$2 \vee 3$'', ``$2 \vee 4$'', and ``$3 \vee 4$''~\cite{Spekkens2007}.
Here, $\vee$ denotes the disjunction or ``OR'' operator.
These states can be represented by column vectors
\begin{equation}
\begin{split}
&\eta_5=
\begin{pmatrix}
\frac{1}{2}\\ 
\frac{1}{2}\\ 
0\\ 
0
\end{pmatrix},
\eta_6=
\begin{pmatrix}
0\\ 
0\\
\frac{1}{2}\\ 
\frac{1}{2}
\end{pmatrix},
\eta_7=
\begin{pmatrix}
\frac{1}{2}\\ 
0\\ 
\frac{1}{2}\\ 
0
\end{pmatrix},
\eta_8=
\begin{pmatrix}
0\\
\frac{1}{2}\\ 
0\\ 
\frac{1}{2}
\end{pmatrix},\\
&\eta_9=
\begin{pmatrix}
0\\
\frac{1}{2}\\ 
\frac{1}{2}\\ 
0
\end{pmatrix},
\eta_{10}=
\begin{pmatrix}
\frac{1}{2}\\
0\\ 
0\\ 
\frac{1}{2}
\end{pmatrix},
\end{split}
\end{equation}
respectively, which form an octahedron that is circumscribed by the GPT's state space\footnote{It is worth emphasizing that, despite the representation given here, Spekkens' toy theory does not allow for all possible convex combinations of its pure states~\cite{Spekkens2007}. However, this fact is irrelevant to our analysis.}; see Fig.~\ref{fig:Spekkens}.
The set of reproducible and nonrefinable PVVMs also reduces to $\{1 \vee 2,3 \vee 4\}$, $\{1 \vee 3,2 \vee 4\}$, and $\{1 \vee 4,2 \vee 3\}$, that can be represented by
\begin{equation}
\begin{split}
&\zeta_5=
\begin{pmatrix}
1\\ 
1\\ 
0\\ 
0
\end{pmatrix}\trs,
\zeta_6=
\begin{pmatrix}
0\\ 
0\\
1\\ 
1
\end{pmatrix}\trs,
\zeta_7=
\begin{pmatrix}
1\\ 
0\\ 
1\\ 
0
\end{pmatrix}\trs,
\zeta_8=
\begin{pmatrix}
0\\
1\\ 
0\\ 
1
\end{pmatrix}\trs,\\
&\zeta_9=
\begin{pmatrix}
0\\
1\\ 
1\\ 
0
\end{pmatrix}\trs,
\zeta_{10}=
\begin{pmatrix}
1\\
0\\ 
0\\ 
1
\end{pmatrix}\trs,
\end{split}
\end{equation}
respectively, which also form an octahedron inside the GPT's effect space; see Fig.~\ref{fig:Spekkens}.
By the restriction enacted, the effect and state spaces of the subGPT do not satisfy the no-restriction hypothesis anymore.
It is clear that $\dim\V\s{Spekkens}=4$ equals the dimensionality of the container GPT.
A second way to check this is by considering the {\it probability table} of the subGPT, given as
\begin{equation}\label{eq:TSpekkens}
T\s{Spekkens}=
\begin{tabular}{l|ccccccc}
 &  E(5)   &  E(6)   &   E(7)  &   E(8)  & E(9) & E(10)  \\ \hline
$\pp_5$ & 1   & 0   & 1/2 & 1/2 & 1/2 & 1/2 &  \\
$\pp_6$ & 0   & 1   & 1/2 & 1/2 & 1/2 & 1/2 & \\
$\pp_7$ & 1/2 & 1/2 & 1   & 0   & 1/2 & 1/2 & \\
$\pp_8$ & 1/2 & 1/2 & 0   & 1   & 1/2 & 1/2 & \\
$\pp_9$ & 1/2 & 1/2 & 1/2 & 1/2 & 1 & 0 & \\
$\pp_{10}$ & 1/2 & 1/2 & 1/2 & 1/2 & 0 & 1 &
\end{tabular},
\end{equation}
and checking its rank.
In this case, $\rank T\s{Spekkens}=4$.
Given a probability table for sets of pure preparations and nonrefinable measurements, it is known that any model for that table must satisfy $\dim\V\s{model}\geqslant \rank T$~\cite{Janotta2014}, where for the Spekkens' toy model the equality can be satisfied.
Consequently, by Theorem~\ref{th:GPTnecsuf}, Spekkens' subGPT is a proper ontologically noncontextual subtheory.

\subsection{Stabilizer rebit Theory}\label{subapp:rebit}

Our second example is closely related to but significantly different from the previous one.
The subtheory of stabilizer rebit is a model in which the nonrefinable effects and pure states of a qubit system is restricted to $E(1),\pp_1\mapsto\ketbra{0}$, $E(2),\pp_2\mapsto\ketbra{1}$, $E(3),\pp_3\mapsto\ketbra{+}$, $E(4),\pp_4\mapsto\ketbra{-}$.
These produce a table of probabilities for {\it all in principle possible} prepare and measure experiments on a rebit given by,
\begin{equation}\label{eq:Trebit}
T\s{rebit}=
\begin{tabular}{l|ccccc}
 &  E(1)   &  E(2)   &   E(3)  &   E(4)  &  \\ \hline
$\pp_1$ & 1   & 0   & 1/2 & 1/2 &  \\
$\pp_2$ & 0   & 1   & 1/2 & 1/2 &  \\
$\pp_3$ & 1/2 & 1/2 & 1   & 0   &  \\
$\pp_4$ & 1/2 & 1/2 & 0   & 1   & 
\end{tabular}.
\end{equation}
It is evident that $T\s{rebit}$ is a subtable of $T\s{Spekkens}$ in Eq.~\eqref{eq:TSpekkens}.

Suppose that we want to construct a GPT for this table.
The first question is what the dimensionality of the employed vector space should be.
To answer this question, we note that $\rank T\s{rebit}=3$ and thus, regardless of the model constructed being a GPT or an ontological model, $\dim\V\s{model}\geq 3$.

A GPT on a three dimensional space reproducing $T_1$ can be given by the states
\begin{equation}\label{eq:GPTstateV3}
s_1=
\begin{pmatrix}
\frac{1}{2}\\ 
0\\ 
\frac{1}{2}
\end{pmatrix},
s_2=
\begin{pmatrix}
-\frac{1}{2}\\ 
0\\
\frac{1}{2}
\end{pmatrix},
s_3=
\begin{pmatrix}
0\\ 
\frac{1}{2}\\ 
\frac{1}{2}
\end{pmatrix},
s_4=
\begin{pmatrix}
0\\
-\frac{1}{2}\\ 
\frac{1}{2}
\end{pmatrix},
\end{equation}
and the effects
\begin{equation}\label{eq:GPTeffectsV3}
e_1=
\begin{pmatrix}
1\\ 
0\\ 
1
\end{pmatrix}\trs,
e_2=
\begin{pmatrix}
-1\\ 
0\\
1
\end{pmatrix}\trs,
e_3=
\begin{pmatrix}
0\\ 
1\\ 
1
\end{pmatrix}\trs,
e_4=
\begin{pmatrix}
0\\
-1\\ 
1
\end{pmatrix}\trs,
\end{equation}
with the unit element $e_U=( 0, 0, 2)$.

We note that the given GPT does not satisfy the no-restriction hypothesis, i.e. it is a subGPT. 
One GPT containing this subGPT, for instance, is given by the state space generated by the vectors
\begin{equation}\label{eq:GPTstateV3pure}
s_5=
\begin{pmatrix}
\frac{1}{2}\\ 
\frac{1}{2}\\ 
\frac{1}{2}
\end{pmatrix},
s_6=
\begin{pmatrix}
-\frac{1}{2}\\ 
\frac{1}{2}\\
\frac{1}{2}
\end{pmatrix},
s_7=
\begin{pmatrix}
\frac{1}{2}\\ 
-\frac{1}{2}\\ 
\frac{1}{2}
\end{pmatrix},
s_8=
\begin{pmatrix}
-\frac{1}{2}\\
-\frac{1}{2}\\ 
\frac{1}{2}
\end{pmatrix}.
\end{equation}
These states form the extreme points of the dual cone to the effect space defined by $e_1$, $e_2$, $e_3$, and $e_4$ with a positive inner product with all the given effects and thus, if the no-restriction hypothesis holds, they must be legitimate states.
It is clear that the states  $s_1$,$s_2$,$s_3$, and $s_4$ can be written as convex combinations of $s_5$,$s_6$,$s_7$, and $s_8$. 
For instance, $s_1=(s_5+s_7)/2$.
From Theorem~\ref{th:GPTNCOM} follows that this GPT is ontologically contextual.

Given the four pure states and the four nonrefinable effects of table $T\s{rebit}$ in Eq.~\eqref{eq:Trebit}, no ontological model that respects the broad noncontextulity and reproduces the desired probabilities exists for $\card\Upsilon{=}\dim\V\s{ontic}{=}3$.
Hence, the subGPT is ontologically contextual for $\card\Upsilon{=}\dim\V\s{ontic}{=}3$.
We now show that increasing dimensionality does not resolve this issue.

Quite easily, it is possible to construct an ontological model for the above subGPT by enlarging the ontic space such that $\card\Upsilon{=}\dim\V\s{ontic}{=}4$.
We can represent preparations and measurement effects as ontic states and indicator functions
\begin{equation}\label{eq:onticstateV4}
\eta_5=
\begin{pmatrix}
1\\ 
0\\ 
0\\ 
0
\end{pmatrix},
\eta_6=
\begin{pmatrix}
0\\ 
1\\
0\\ 
0
\end{pmatrix},
\eta_7=
\begin{pmatrix}
0\\ 
0\\ 
1\\ 
0
\end{pmatrix},
\eta_8=
\begin{pmatrix}
0\\
0\\ 
0\\ 
1
\end{pmatrix}
\end{equation}
and
\begin{equation}\label{eq:onticeffectsV4}
\zeta_5=
\begin{pmatrix}
1\\ 
0\\ 
0\\ 
0
\end{pmatrix}\trs,
\zeta_6=
\begin{pmatrix}
0\\ 
1\\
0\\ 
0
\end{pmatrix}\trs,
\zeta_7=
\begin{pmatrix}
0\\ 
0\\ 
1\\ 
0
\end{pmatrix}\trs,
\zeta_8=
\begin{pmatrix}
0\\
0\\ 
0\\ 
1
\end{pmatrix}\trs,
\end{equation}
respectively.
This is exactly the GPT containing Spekkens' toy theory.
There are various ways in which we can construct a subGPT reproducing $T\s{rebit}$, one of which is via the states and indicator functions
\begin{equation}
\eta_1=
\begin{pmatrix}
\frac{1}{2}\\ 
\frac{1}{2}\\ 
0\\ 
0
\end{pmatrix},
\eta_2=
\begin{pmatrix}
0\\ 
0\\
\frac{1}{2}\\ 
\frac{1}{2}
\end{pmatrix},
\eta_3=
\begin{pmatrix}
\frac{1}{2}\\ 
0\\ 
\frac{1}{2}\\ 
0
\end{pmatrix},
\eta_4=
\begin{pmatrix}
0\\
\frac{1}{2}\\ 
0\\ 
\frac{1}{2}
\end{pmatrix}
\end{equation}
and
\begin{equation}\label{eq:rebiteffectsV4}
\zeta_1=
\begin{pmatrix}
1\\ 
1\\ 
0\\ 
0
\end{pmatrix}\trs,
\zeta_2=
\begin{pmatrix}
0\\ 
0\\
1\\ 
1
\end{pmatrix}\trs,
\zeta_3=
\begin{pmatrix}
1\\ 
0\\ 
1\\ 
0
\end{pmatrix}\trs,
\zeta_4=
\begin{pmatrix}
0\\
1\\ 
0\\ 
1
\end{pmatrix}\trs.
\end{equation}

The most important point here is that, despite being an ontological model, this model does not satisfy broad noncontextuality when noting that we are operationally restricted to the statistics given by the subGPT.
In other words, assuming the subGPT $\T\s{rebit}$ (or in general, any GPT) implicitly implies that no preparation and measurement beyond what is predicted by the subGPT exist.
As a result, we can readily verify that the vector
\begin{equation}
\eta^\star=\begin{pmatrix}
\frac{1}{2}\\
0\\ 
0\\ 
\frac{1}{2}
\end{pmatrix},
\end{equation}
produces the same statistics as the maximally mixed rebit state
\begin{equation}
\eta\s{mm}=\begin{pmatrix}
\frac{1}{4}\\
\frac{1}{4}\\ 
\frac{1}{4}\\ 
\frac{1}{4}
\end{pmatrix},
\end{equation}
for every {\it physically legitimate} ontic effect generated by $\{\zeta_5,\zeta_6,\zeta_7,\zeta_8\}$ of Eq.~\eqref{eq:rebiteffectsV4}.

Note that, $\eta^\star$ does not belong to the state space of the stabilizer rebit subGPT $\St\s{rebit}$ but it is an ontic state of the full ontological model, i.e. it is a convex combination of states in Eq.~\eqref{eq:onticstateV4}.
As such, given the preparation procedure $\pp\s{mm}$ that generates the table
\begin{equation}\label{eq:Tmm}
T\s{mm}=
\begin{tabular}{l|ccccc}
 &  E(1)   &  E(2)   &   E(3)  &   E(4)  &  \\ \hline
$\pp\s{mm}$ & 1/2   & 1/2   & 1/2 & 1/2 &  
\end{tabular},
\end{equation}
we have two possible ontic assignments:
\begin{equation}\label{eq:1to2}
\pp\s{mm}\mapsto \eta\s{mm} \quad \text{or} \quad \pp\s{mm}\mapsto \eta^\star.
\end{equation}
Therefore, eventhough $\eta\s{mm}$ and $\eta^\star$ are distinct ontic functions with regards to the full ontic measurements of Eq.~\eqref{eq:onticeffectsV4}, by the implicit assumption of the stabilizer rebit subGPT, it is operationally impossible to distinguish them.
As a result, Eq.~\eqref{eq:1to2} is in clear contradiction with broad noncotextuality assumption for NCOMs and the fact that each preparation equivalence class must be represented by a {\it unique} ontic state that is completely determined by the observable statistics. 

It is worth pointing out that similar arguments can also be made by exchanging the roles of ontic states and indicator functions. 

\section{Is there a conflict?} \label{app:conflict}

An interesting lesson we learn from the example of stabilizer rebit subtheory is that restricting classical theories (in this case the ontologically noncontextual GPT given by Eqs.~\eqref{eq:onticstateV4} and~\eqref{eq:onticeffectsV4}) in certain ways may also give rise to ontologically contextual subtheories.
This will face us with the dilemma of choosing between (i) assuming all subtheories of classical theories are classical and thus ontological contextuality is not a purely nonclassical effect, or (ii) ontological noncontextuality is a purely nonclassical effect and thus subtheories of classical theories can also be nonclassical.
Our personal preference is the latter, because the assumption of (i) cannot be justified.
In particular, there is no reason to call a weird-looking world wherein our empirical observations are restricted in such a way that certain statistics have multiple undeterminable ontic explanations classical, noting that any such explanation is different from our current classical perception of the world. 

One might also be concerned that the claim ``stabilizer rebit subGPT is ontologically contextual" conflicts with other results, e.g. Refs.~\cite{Schmid2018} and ~\cite{Schmid2018PRA}, or that this subGPT can be classically {\it simulated}.
We, however, emphasize that there is no conflicts with either of these results for the following reasons.

In Refs.~\cite{Schmid2018,Schmid2018PRA}, the idea is to build an NCOM for a given {\it table of probabilities} rather than a subGPT.
There is a significant difference between the two, namely that, once a subGPT (like the stabilizer rebit theory) is given it is assumed that
\begin{quote}
no observable statistics exist beyond those predicted by the (sub)theory.
\end{quote}
Therefore, if an NCOM exists in larger spaces, the extra statistics predicted by the NCOM is empirically unobservable leading to the previously discussed contradiction. 
In sharp contrast, given merely a set of experimental data there is {\it no} assumption that the given data is {\it all one can in principle measure for the physical system}.
Hence, once an NCOM for the data is built (possibly in larger dimensions) it is implicit that the extra statistics predicted by the model but not included in the given table are indeed empirically observable.
Hence, the difference between the two scenarios is in assuming or not assuming empirical observability of the statistics predicted by the NCOM beyond the given table or the subGPT, respectively. 

A similar argument holds when we speak of classical simulatibility, wherein, again, there is no assumption of physical possibility or impossibility of certain statistics.
Therefore, as an example, the stabilizer rebit subGPT being ontologically contextual does {\it not} imply it being classically nonsimulatable.
This highlights the conceptual difference between ``classicality" and ``classical simulatibility".
Our position of classicality identified by ontological noncontextuality requires, in addition to classical simulatibility, that the (sub)GPT is not underdetermined in the sense that there are no {\it in principle inaccessible} variables present at the ontological level.
The latter requirement follows the spirit of Leibniz's methodological principle, as Spekkens~\cite{Spekkens2019} puts it,
\begin{quote}
``If an ontological theory implies the existence of two scenarios that are empirically indistinguishable in principle but ontologically distinct (where both the indistinguishability and distinctness are evaluated by the lights of the theory in question), then the ontological theory should be rejected and replaced with one relative to which the two scenarios are ontologically identical.''
\end{quote}
\noindent The principle thus asks for ontologically identical descriptions of facts that cannot be empirically distinguished. 
It is clear that, in the example of rebit stabilizer subGPT, replacing $\eta\s{mm}$ with $\eta^\star$ results in two ontologically distinct accounts for the same subGPT (i.e. empirical) description.
Hence, the rebit stabilizer subGPT does not allow for an ontological model satisfying the Leibniz's methodological principle and, by our definition, it is a nonclassical subGPT.

\section{Proof of Theorem~\ref{th:GPTRes}} \label{app:GPTRes}

That (i) if and only if (ii) follows from the definition of nonclassical resources given above.
That (i) if and only if (iii) follows from Theorem~\ref{th:GPTNCOM} and that $\EE\s{nr}$ (pure states) forms a complete basis for $\V$.
That (iii) if and only if (iv) follows from the assumption that $\EE\s{nr}$ (pure states) form a complete basis for $\V$, hence adding an extra nonrefinable extreme point overcompletes it if $\V^\star{=}\V$. Conversely, it is trivial that if an overcompleting element lies within $\EE$, it is either refinable or a convex combination of the extreme points.
\qed

\bibliographystyle{apsrev4-1new}
\bibliography{ContextualityInGPTs}

\end{document}